# Smart business networks and business genetics

## with a high tech communications supplier selection industry case

L-F Pau, Prof. Copenhagen Business school and Rotterdam school of management

**Abstract:**
Despite the emergence of event driven business process management, smart business networks, social networks, etc. as important research areas in management, for all the attractiveness of these concepts, two major challenges remain around their design and the partner selection rules while learning from interaction events. While smart business networks should provide advantages due to the quick connect of business partners for selected functions in a process common to several parties, literature does not provide constructive methods whereby the selection of temporary partners and functions can be done. Most discussions only rely solely on human judgment. This paper introduces both computational geometry, and genetic programming, as systematic methods whereby to identify, characterize, and then display on a continuing basis from event monitoring such possible partnerships; such techniques also allow to plan for their effect on the organizations and thus to carry out selection. The two methods are being put in the context of emergence theory. Tessellations address the identification and categorization issues; business maps address the display and monitoring challenge with the use of Voronoï diagrams. Cellular automata mimicking living bodies, with genetic algorithms of which parameters are estimated by learning, address the selection and effect issues. To illustrate the approach, some experimental results from the sourcing function in a high tech industry, are discussed; they address the case of how to determine the selection process for a systems integrator to set up joint ventures with smaller technology suppliers.

## 0. INTRODUCTION

The management concepts around “smart business networks” have appeal to general management as a structuring notion, and to information management as a means to highlight how communications and information can structure organizations and not just processes. However, these concepts and related management research fail to offer the needed methodologies and tools whereby the underlying forces (creative , disruptive ,or regenerative) can be identified ; they do not either give means to elicit ,in view of discovery of partners, the business and organizational rules whereby such smart business networks actually can exhibit any smartness (as defined and discussed in (Vervest, Heck, Preiss, Pau(2005)).The difficulty lies in large parts in the fact that general management, information management, and business process management being in the philosophical sense "reductionist" approaches with a tendency towards static mechanisms , they cannot tackle the complexity and volatile behaviours encountered in what are considered as being real live cases of smart business networks . The notion of "business operating system" is one such example of a very limited reductionist if not even normative view of such networks.

This paper goes well outside avenues normally considered in general management and information management, by taking inspiration from emergence theory in philosophy on one hand, from genetics, cellular automata and computational geometry on the other hand, to identify methodologies and tools helpful in eliciting the business and organizational rules whereby smart business networks can eventually exhibit common forms of smartness.

Also, the paper is discussing an implementation case, by reviewing the proposed approach for evaluation purposes. While details cannot be given due to their very real and time critical nature, this part of the paper shows how a large systems integrator in the high tech field can decide to initiate and terminate technology agreements (or technology joint venture agreements) with smaller technology suppliers.

By addressing such issues in this way, this paper points at business genetics as an interdisciplinary research area with direct business implications, from technology management, to mergers and acquisitions, to business process strategy. By business genetics is here meant the application and adaptation of genetic processes from biology and cellular automata, to business relations and organizations.

## 1. CRITICAL REVIEW OF SMART BUSINESS NETWORK NOTIONS

The initial concepts in smart business networks were presented in (Vervest, Heck, Preiss, Pau (2004); Vervest, Heck, Preiss, Pau (2005)) with as core notions:

1. Agility (Pal et al, 2005),
2. No definite lasting commitments (uni-or multilateral),
3. Process specific, and
4. The use of communication and signalling networks as command / synchronization lines.

A smart business network does not have to rely on mutual equivalence structures (Weick (1979)) defined as implicit contracts between people that can be built and sustained without knowing the motives of another, and without sharing goals. A smart business network may have to use network management rules, or so-called "business operating systems", provided they preserve all the evolutionary aspects (, (Bahrami (2005), Klein et al (2005), Pau (2005)), but may also develop without such rules up to a certain level of complexity to avoid the reductionist implications of such rules.

Smart business networks are not informal networks as found in most social groups and even some industry sectors .Although they retain the innovative dimension of informal networks (Ehin (2005)) they have designed organizational temporary processes as opposed to those derived from situation dynamics and hidden elements only .

Smart business networks are also a departure from knowledge sharing concepts (or networks of practices) where parties have to agree on goals and behave consistently ;the difference resides in the explicit organizational existence and recognition of different interests , preferences, and goals across the smart business network .In order to achieve some of these ends, initiating parties have to initiate actions towards others by which they create mutual commitment and interlocked behaviours ,to collectively pursue diverse ends through selected common means .Once people are engaged in selected mutual commitments a subtle shift takes place from diverse to common ends .Diverse ends remain but they become subordinated to an emerging set of shared ends. We characterize this evolution by the dynamics of the ratio of shared goals to total goal set cardinality ,for each party ; if N(i)= Card ( Goals(i)) is the cardinality of all goals of party (i) , the ratio is initially 1/N(i) for one Card(i)=1 party ,to reach a value q , larger or equal to 1/N(i), in a smart business network involving Card(i) = p parties ; much of the dynamics in the smart business network can be seen from the time-dependent evolution graph of the pair (p,q) called smart business network consistence graph . This graph is comparable with the genetic signature fit between living cells in biology and genetic engineering.

Smart business networks are a departure as well from business process management systems (Chang (2005)) as BPMS standards and service oriented architectures (SOA) rely on identifying the full set of capabilities , data integration , messaging based integration ,and software component based integration steps needed to execute a specified process . Business process management relies on decomposition of the process and task allocation, not on the mutual fit of the parties for other tasks or knowledge.

Smart business networks however depend on the existence of reciprocity. The motivation for people and organizations to contribute to an online connected group of people or organizations who do not or hardly know each other is still an area of research. Wellman and Gulia (1999) point to different types of explanations for such motivations .The first refers to the fact that online contributions are a means of expressing one's identity : helping others might increase self-esteem , reputation , respect from others ,etc ...The second one is generalized reciprocity and organizational citizenship .McLure Wasko and Faraj (2000) state that sharing knowledge and helping others is "the right thing to do" and that people also have the desire to advance the community as a whole .Members in a smart business network may not expect to be reciprocated by the same person or organization with whom they share knowledge or transactions ,but they do expect to receive future help or transactions from someone in the network .Also contributions being via online networks are at low participation and switching costs ,so there is the constant risk of network failure if active knowledge or transaction producers withdraw . Thus smart business networks do not have three of the characteristics of social networks, which are: ongoing interaction, identity persistence, and knowledge of the previous interactions (Kollock, 1999). Smart business networks are as fragile as minimal social situations in emergent social networks.

This brings focus (for the second time after the business network consistence graph) on a characteristic of smart business network ,that is their similarity with genetic processes where the networking effect results from mutual perceived forces of attraction (or repulsion) and on evolutionary birth-life-death processes .For "A" , a user or economic agent, it is not only important to know if "A" prefers to use a particular process ,but also if other agents "B" and "C" have similar preference and expectation values for the same process before they link up .In such a case , "A" may choose to access "C" via "B", if "A" feels more attracted by "B" and if "B" and "C" have a bilateral agreement ,and not necessarily set up a network involving jointly "A","B" and "C" . The implication of this argument is that the design of a smart business network must not only reflect the needs of individual members, but more importantly the social triadic relationships in the emerging networks and how they evolve over time (Wenger et al (2002)). If the perceived forces of attraction grow, maybe "A" and "B" will merge. If the perceived forces of repulsion grow, "A" or "C" may quit the network and it is destroyed. As to the evolution over time, two types of alternative forces may apply: either genetic conquer or divide principles, or statistical dynamics with random walks; in this last case there are relations between how far nodes in a smart business network may jiggle over time, the number and size of the nodes, and the "viscosity" and diffusion coefficient in the environment. This allows stating the usefulness of Monte Carlo simulation to analyze these effects at a statistical level.

This leads directly to proposing a representation of smart business networks, and a design method, relying on the connectivity in genetically evolving networks, with topologies found both in the business relations space as well as in their communications network topologies. The following Section 2 will discuss how this representation is routed in philosophical theories departing from those used in traditional general management.

## 2. SMART BUSINESS NETWORKS AND EMERGENCE THEORY

Emergence theory (Sober (2004 ),Holland (1999, 2000), Kim (1999) ) is a recent line of research in philosophy and physics aiming at filling some gaps found in analytical theories of evolution .It also claims that the world is not made of assemblies of particles, components or processes interacting with each other, but instead of a large variety of objects and processes each having singular definitions and obeying each to their own rules .In this theory , the whole should be more than the sum of its parts It is said that a property or a process are "emergent" at a given organizational level if, although in principle reducible to the properties of its constituents at a lower level (Glymour(1970)),it's sudden appearance seems impossible to predict a priori from knowledge of these properties .

A special class of emerging organizations , are cellular automata which create higher level organizations when some conditions are met between their constituents ; such organizations can experience chaos as well as order at times , can dissipate properties , can oscillate in synchronism ,etc ...Attraction (also called attractive) nodes are fixed or periodic or chaotic configurations towards which the evolution may go although in different ways (Berlekamp (2001) , Heudin ( 1994 ), Heudin (1998) ) .Each is nevertheless the subject to internal as well as system-wide effects which cause the change and evolution . As an example, a bifurcation or split happens in an attractive node when a system parameter reaches a critical value creating dis-homogeneity in the whole system.

The illustrative set of rules by John Conway called "game of life" (See :www.virtual-worlds.net) is:

1) If a node in state "1" is surrounded by two or three nodes in state "1», then it keeps its state

2)if a node is in state "0" and is surrounded by three nodes in state "1" ,then it changes to state "1"

3) In all other cases, the node switches to state "0"

Starting from a random configuration of nodes in either state "0" or "1», after some tens of iterations the population of nodes in state "1" dwindles fast, being replaced by new constellations of nodes with changing shapes but with some stability; some of these constellations emerge in different orientations or positions. In Section 4, a richer formalism with associated rules will be presented which is closer to the needs of business genetics.

Cellular automata governed by different sets of rules are categorized into:

-Class I: evolution towards fixed stable configurations irrespective of the initial configuration

-Class II: evolution towards stable periodic configurations, after some iterations

-Class III: evolution towards a succession of chaotic configurations, which nevertheless share a same property (e.g. the proportion of nodes in state "1")

-Class IV: evolution towards the emergence of long transitory configurations (blocks, beehives, blinkers, gliders, etc ...) with a large diversity which seem to interact with one another.

Experimentation with cellular automata allows to study self-organization phenomena, and above all, in specific domains, to elicit the set of rules which lead to different types of organizations over time .Have for example been already researched the evolution from collective oscillations to collective chaos (Nakagawa (1994)), the emergence of collective behaviour in large chaotic systems (Chate (1998)), and synchronization (Pikovsky (2001)).

The full relevance of cellular automata for smart business networks, including their dynamics, will emerge in Section 3, 4.2 and 5, but it can already be conjectured here that they offer a valuable simple formalism to analyze:

-business relations represent forces and energy, some measurable and others not, which reshape business partnerships

-the state of the business activities, instead of being just reduced to two states, can be partitioned into either finite states, or finite classes of risks

-while business deals may drive the initial engagements , a very interesting issue are the set of rules whereby they are modified or cancelled in view of sector/economy wide forces and reorganization .

But what is missing is now to relate such automata, and the emergent behaviours, with business analysis and possible representation tools helping first illustrate the smart business network dynamics, and next elicit the applicable forces or rules.

## 3. SMART BUSINESS MAPS

A number of authors ,even prior to the “smart business network” concept and theory emergence in 2004 , have mapped out snapshots at one point in time of business relations; in the simplest cases they used attributed graphs (with attributes being a few one-one-one relationship measures such as purchases/sales) ,or in more complex cases they used multivariate causal analyses such as correspondence analysis between large sets of companies and financial results over time .Hierarchical data structures such as quad trees suffer from the same problems as attributed graphs. All these approaches (attributed graphs, quad trees, causal graphs) suffer from a static approach to the information as well as to the results .Attributed graphs furthermore give , to the contrary of correspondence analysis , no revelations of causal factors or business preferences explaining some key decisions . By working on time indexed information, correspondence analysis however can reveal graphically some evolutions and the drivers therefore, as exemplified in the analysis of Danish bank’s balance sheets over time (Pau (1977)). It results from these early lessons that multiple dimensions must be taken into account, but that there is a trade-off in the complexity of the analysis as well more importantly in its visualization for intelligibility.

We will use below the notion of “smart business map” to designate a visualization of the forces of attraction and repulsion (see Sections 1 and 2), each represented by one graph dimension, between several potential parties in a possible smart business network; a simple illustration is given in Figure 1.

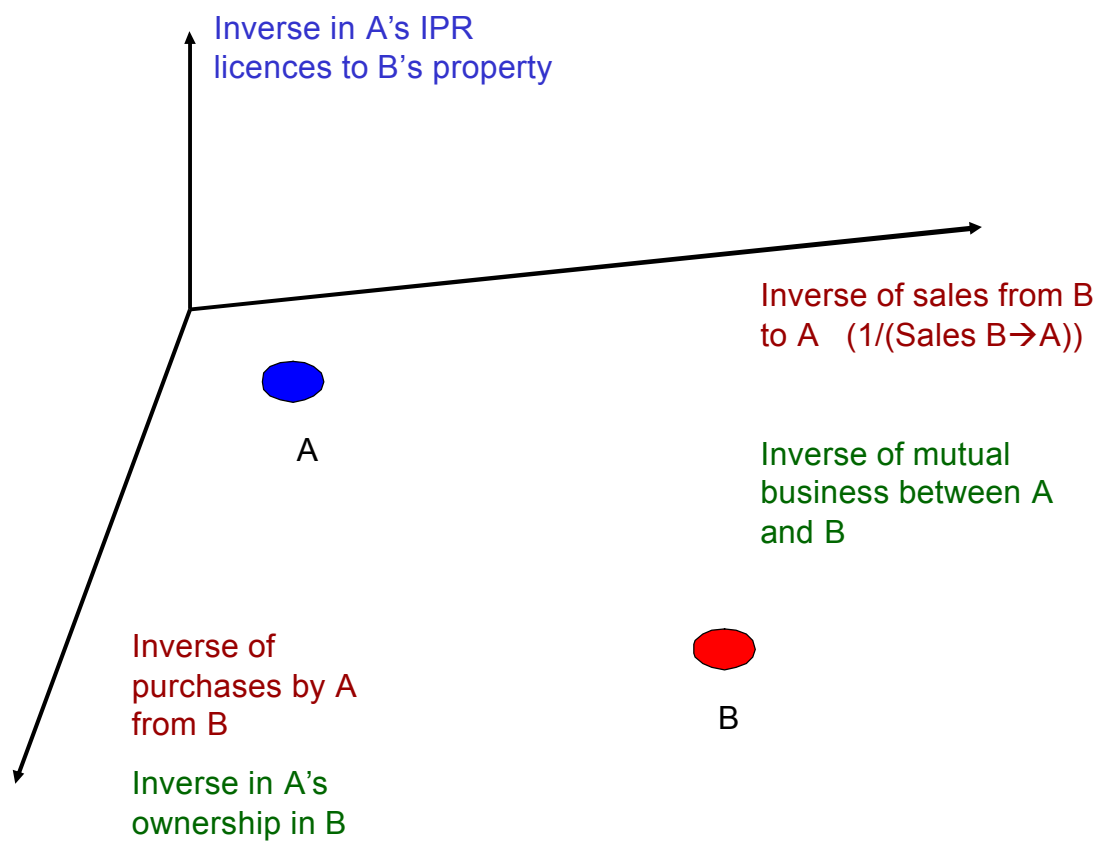

Figure 1: Examples of dimensions in a smart business map

In a smart business map, examples of dimensions are:
Example 1 (2 dimensions): x-deviation from B to A representing the inverse of sales from B to A, while the y deviation from B to A represents the inverse of purchases by A from B
Example 2 (2 dimensions): x deviation from B to A represents the inverse of the total business notions: volume between the two parties, while the y deviation from B to A is the inverse of A's ownership share in B
Example 3 (3 dimensions): add to Example 3 the z-deviation from B to A as the inverse in licences (number or license payments) by A to B's intellectual property rights
The notion and use of smart business maps allows in the next Section 4 to introduce computational geometry and its powerful rules.

# 4. VORONOÏ DIAGRAMS AND COORDINATION

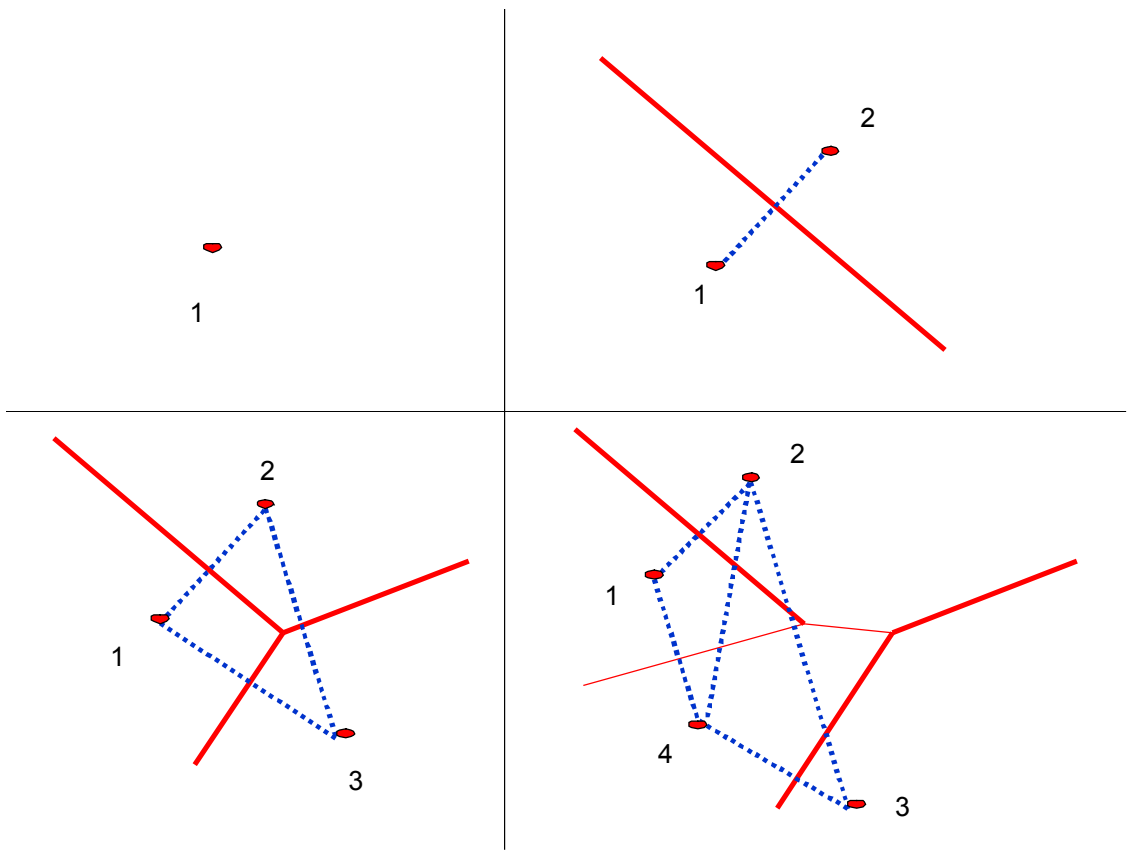

Figure 2: Basics in Voronoï diagrams and tessellations; the red points are the nodes; the red lines are the equi-distance Voronoï lines; the figure illustrates the insertion of a second, a third and a fourth node

## 41. VORONOÏ TESSELATION

Consider a set of objects (points) in the plane .Each of these objects is to be considered to have a sphere of influence, defined as the region which is closer to that object than to any other object .The result of this zoning activity is to partition the plane into a set of polygonal regions, each region associated with a particular object .For points in the plane these polygonal regions can be shown to be convex polygons .The result of this process is referred to as a Voronoï tessellation.
While the mathematical definition is straightforward it must be emphasized that Voronoï diagrams are not at all abstract entities. This approach may be created e.g. in physics by magnetic fields, but does apply also to business processes if e.g. the distance measure is the level of business exchanges between two parties represented each by a point .Thus Voronoï diagrams are closely related to real physical or business processes which simplifies both the visualisation of the technique and the potential for the modelling of these processes.
Considerable research has been dedicated to studying Voronoï diagrams. While theoretical algorithms are the particular speciality of the field of computational geometry, the applications in business process management have not been explored. The efficient construction of point Voronoï diagrams in the Euclidean space has been well known for some years ,but other particular Voronoï diagrams (using other metrics, furthest point Voronoï techniques , cases with boundaries ,etc ) are still research issues (Preparata & Shamos(1985)).As a general statement, coordinates do not of themselves produce relationships , that is :graph theoretical structures relating objects in space .This is partly due to the fact that the two branches of mathematics involved have very little overlap in these problems . Graph theoretic techniques require that relationships (adjacency relationships in particular) be previously defined , while the straightforward definition of coordinates provide no information of itself about the linkage between points and objects in space .It is here suggested that the use of a Voronoï generating process may simplify the transition from coordinate based information to graph theoretic adjacency based structures .

42. BUSINESS GENETICS VS COMPUTATIONAL GEOMETRY

The above approach has various characteristics, which include the use of "split, grow, merge» as in cellular automata and even more explicitly in genetic algorithms, or "divide and conquer" methods to obtain the most efficient construction techniques. While the use of "divide and conquer" techniques implies the construction of the diagram for the whole data set at one time (read: the whole business network), the genetic techniques allow fundamentally the updating of the data set in the process of the application (read: the business growth or demise). The concept of biologically inspired network has been considered in engineering, but so far not in management
(Chlamtac,Carreras,Woesner,2005)
(Suzuki,Suda, 2005).But the derived data structures are interesting in that they define gene footprints by the combinations of (value, timing information, information source , other services, and they consider that the service is the gene organism itself.

Business genetics should look beyond a game a survival in that reproduction and evolution should be driven by general evolution rules, such as those from Section 4.3. Also , the determination of the exchange of information is driven by “fitness” which is the correspondence of the organism’s genetic information footprint with the environment .Information is exchanged as needed , locally , between mating organisms ,like in smart business networks .

43. IMPLEMENTATION AND STORAGE OF VORONOÏ POLYGONS

As previously mentioned, the computational geometry approach to point Voronoï diagram generation is based on "divide and conquer" methodology, whereby the whole data set is "inserted " at once .This approach is based on the assumption that the input data will require significant small scale adjustments before it is in its final form, thus the emphasis must be on local operations for the insertion and deletion of individual points and line segments.

As the guiding principle is that adjacency of objects is defined by the adjacency of their Voronoï regions ,and these regions occupy all the available two dimensional space (read :all levels of business transactions up to the limits set by the convex regions) ;but the first point inserted "owns" the available universe ,just like an innovative enterprise with a new product/service does .The second point is generated by the first point splitting in a cell-like fashion ,or by a selection of the first business partner ,the new point then moving to its final location as the business first is created. The universe now has two Voronoï cells, with a linear boundary between them .Subsequent points are formed either by cellular subdivision of a suitable nearby point, followed by local movement, or by the selection of a very remote outsider for a closer relation. The “*split”* operation described previously is an action that can be considered to be a division of a general polygon into two, with the generation of two adjacent dual triangles that specify the new adjacency relationships formed, or the insertion of a new node into a generalized linked list.

The "*delete*" operation is the reverse activity: the moving of the point to be deleted to a nearby one, followed by the "merge" of the two adjacent polygons .This again may be thought of as the deletion of their now redundant common boundary, the deletion of the two unnecessary dual triangles, or the removal of a node from the generalized linked list.

We thus have a cellular life cycle: birth (split, acquire), life (move, grow) and death (shrink, merge, divest) .This leads to the study of smart business network dynamics.

## 5. THE SMART BUSINESS NETWORK DYNAMICS

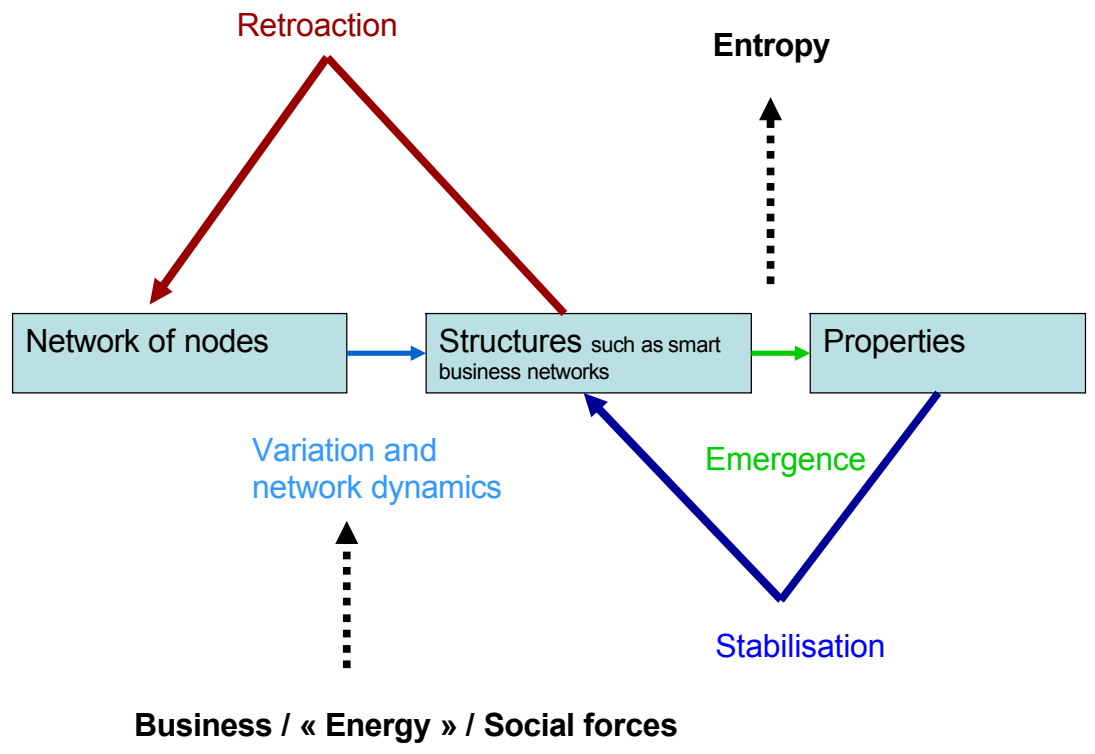

Figure 3: The business drivers and rules in a smart business network, as linked to the effects in a genetic system

Section 4 describes birth and death, but what about life and its evolution? This clearly consists of movement, but any movement will change the boundaries between the moving point and its neighbouring points .For Voronoï polygons, the boundaries may readily be re-created if the neighbours are known, but this information is already stored in the dual triangulation determining the ownership to the respective polygons .Thus small movements that do not change the set of neighbours are not of particular concern and represent a stable business network.

But how can we tell if the neighbourhood set needs to change, and how should it be updated? One definition of the Voronoï criterion is that the circumcircles of the dual triangulation must a empty .Expressed in another way ,the centre of each of these circumcircles is the location where three Voronoï polygons meet ,and thus must be equidistant from the three points or objects forming the triangle vertices .If another point falls within this circle , clearly the previous statement is untrue .Thus if the notion of the moving point takes it inside the circumcircle of an "external " triangle (one that does not contain the moving point as a vertex.) the network must be updated .This is achieving by switching the diagonal of the quadrilateral formed by the external triangle and the moving point (making the triangle vertex that was not an immediate neighbour into one that is now),thus re-establishing the Voronoï criterion (Gold, 1978) .The point is then considered to have moved precisely to the intersection of its proposed path with the circumcircle ,and the next step in its travels is determined by re-examining the neighbouring triangles .

A similar process is followed for the case where the moving point leaves the circumcircle formed by triples of its immediately-neighbouring points. Again a switch of two triangles takes place, but now one of these neighbouring points ceases to be an immediate neighbour, and a newly defined triangle is "left behind".

Thus movement of a point occurs as a series of steps or jumps based on the density of neighbouring points. If we move to higher dimensions, the words "triangles" and "circles" should be updated accordingly ,but otherwise the process is general .It is however limited to point objects

The process just described is sufficient to generate a point Voronoï diagram in any metric. This means in particular that in Smart business maps (Section 3) a wide diversity of properties of business relations and organizations can be used simultaneously to analyze an evolution or elicit rules about their convergence towards a smart business network. But which notion allows taking into account the internal diversity in goals, means and processes as discussed in Section 0? This is addressed in Section 6 which shows the direct relevance to this issue of line segment objects as representations of different business divisions within a given smart business network partner.

## 6. LINE SEGMENT OBJECTS AND BUSINESS DIVISIONS

Line segments are defined as the interiors of line segments, excluding their end points with remain point objects in their own right. In business terms the line segment is the continuum of enterprise groups contributing to a given product or service line (or the wished evolution over time between two positions). A star configuration describes a diversity of product lines or business divisions .The end points can form the vertices (*not* edges) of the dual triangulation .This dual triangulation is referred to as a Delaunay triangulation when limited to point vertices ,and a Voronoï adjacency graph (VAG) when including line-segment or other vertex objects .

Since the Voronoï definition of "zones of influence" about each object , is based on which object any particular location is closest to ,it is readily extendable to any type of object ,and the result must be a planar graph of the polygon set

,just as the VAG must also be planar .
If it is desired to create line-segment objects (read: a continuum of enterprise groups, or an evolution path), it is well known that a line is the locus of a moving point, and hence represents all the previous possible positions of the moving point .Thus a line segment is created by performing a split on its starting point which adds two new objects (a new line head plus a trailing line segment connecting the head with the tail throughout) and four new triangles. The head point is moved as before, but no switching out of previous triangles is performed, the trailing line segment retaining all these previous adjacencies. As a result of these actions, line objects may be created within the business space, either to new coordinate locations or else connecting to previously defined points.
The result of this activity is to generate Voronoï diagrams of any combination of points and line segments that may form a business universe under construction, a polygon set, a set of internal activities plus breaks for partner diversification, etc.
It appears that a large class of smart business network analysis problems, as well as their evolution may be handled through the Voronoï polygons and VAG's whose construction is described above. Note that all operations are incremental, ,and there is no definitive or complete data set involved, so editing can take place at any time ,and polygons that fail to close at the design stage may readily be snapped together during a single pass of the VAG (read : loose partnerships are not sustainable) .

## 7. CASE FROM HIGH TECH COMMUNICATIONS SUPPLIER FIELD

The case is a snapshot of the direct implementation of earlier research reports (Pau, 2005a) (Pau, 2005b) by one of the world's top management consultancies, to assist this research author and his team in their industry to cater to a strategic goal, i.e. turn "A", a high tech systems supplier to the service sector, into a systems and service integrator benefiting from the outsourcing trend amongst its service provider customers. The company Flextronics is also reported to have a modeling software called SimFlex that helps customers figure out the best way to outsource.

### 7.1. CASE SPECIFICATION

More precisely, the case is about designing a smart business network around the field support, installation and consulting Division of the company "A", to allow "A" to achieve a significant worldwide market share in network operations amongst its worldwide service provider customers, at a time where these customers change their core business of running networking services into the new core business of interconnecting networks they do not want to operate themselves anymore. This can only succeed if on a global scale, "A" can identify, select, use and sever links to a wide diversity of smaller technology or skills suppliers, many of them only operating in localized markets, or having de facto only one key customer. As the outsourcing opportunities are time-critical, and as "A" wants to leverage its systems know-how (on its own products and selected other one's), financial terms are in effect of secondary importance compared to a rather large number of intangible properties searched for or to be avoided .Very often the track record of the smaller high tech companies may have been with competitors to "A" or with "A"'s own customers without any direct connection to "A". The potential number of partners in the total smart business network is about 500, with on a country or regional basis a minimum of three and maximum of about 15.

### 7.2. DISCUSSION OF THE USE OF BUSINESS GENETICS IN THIS SMART BUSINESS NETWORK DESIGN

The proposed tools for design and mapping described in the previous Sections were found extremely powerful and relevant first because of the shear automated exhaustive handling of all possible configurations, with their evolutions over time (from known tack records into fulfillment horizons on the outsourcing contracts ) .Next , the possibility for "A", with help of the consulting company, to tailor the forces of attraction and repulsion (usually via simple look-up tables expressing real preferences) around mostly intangibles , was a unique advantage . Intangibles considered fell into the broad categories of : skills sets , available staff on short or medium term notice , prior systems/product/tools experience , incentives and penalty conditions, geographical distance of pockets of skills sets to the customer sites , etc …Third, was considered very valuable the ability with simple cellular automata and computational geometry tools (such as those of Table 1) to project the evolution of the smart business network under risk situations and ,even more valuable, under negotiation sessions ; quick decisions could be made to stop early or to entrust a partner with a wider role ,etc … . Last, was judged very favourably by decision makers the use of smart business maps (see Section 3) for their support and involvement.
The drawbacks were the learning time it took for traditional management consultants to adapt to this novel way of thinking ;but actually this time was far less than the time a merchant bank would have taken to tackle the same volume of analysis .The other drawback was the reluctance by some of the 500 possible parties to disclose some intangible characteristics ;but actually this was never a show-stopper as information was readily available by indirect channels such as the references these same companies were citing .
The outcome parameters were KPI's in supplying outsourcing contracts as business networks, and so far the over 10 instances have not lead to any questioning on the methodology, but rather on the changing goals and structure of the service suppliers.
It is thus considered that the methods can be generalized to other domains where in depth qualitative, quantitative and financial analysis of each possible partner is not possible, due to their number or time pressure. The selection rules must be tailored, although the consulting sector may accumulate experience in this design phase.

## 8. CONCLUSION

The initial scope of this paper was to identify methodologies and tools whereby one could elicit and analyze business and organizational rules whereby smart business networks could be designed to exhibit some smartness.
It has been shown that computational geometry, although of a theoretical nature, in relation to genetic and cellular automata; provide many of the basic insights into the feasibility of various algorithms to design business networking /alliance operations between parties having no/little track records of mutual interactions. It has also been shown above that smart business maps (relying on computational geometry, with business metrics as feature space dimensions), help decision making in the design phase, and allow comparing the actual business performance over time amongst a network of business parties. Individual business parties financial accounts and market achievements do not represent sufficient performance indicators across a smart business network.
This design approach though does not make sense when there is in depth track record of interactions, or human knowledge about the same, or "political" biases in the selection process, cases which anyway usually drift towards much less dynamic smart business network structures.
On the theoretical side, the inability of computational geometry algorithm in providing exact line intersections leads to the weakness of not guaranteeing consistent business network topologies, except with considerable care .In such cases cellular automata, especially of classes III and IV, although more demanding numerically, can provide more robust business and organizational rules.
Finally on the practical implementation side, experiments have started to use the proposed tools and design methodology in the case of content distribution networks driven by instantaneous end user requests and profiles .Here parties have no track record of interaction and their number is much larger, besides the issue that payments must be secured under such circumstances sometimes on a real-time per-request basis .Such a case represent the challenge in the real time design of real time smart business networks !

| **Smart business network design rules , using Vo representation** |
|---|
| 1.Polygons are formed from interconnected vertices<br>And edges .In order for a polygon to be topologically complete<br>Pointers must exist between vertices (if defined ) and edges .The resulting regi then be labelled |
| 2. All vertices (nodes) in a polygonal business map can be forced to have a valen an imaginary zero-length edge and splitting the original node. |
| 3. The dual of a modified polygon set is a triangulation, where all polygons are and all vertices have become triangles. The original arbitrary boundaries between replaced by triangle edges representing an adjacency relationship between polygo |
| 4. Triangulations may readily be stored as fixed length records storing the thr adjacent triangles and, if required , the three bounding edge record numbers for ea |
| 5. An alternative to a triangulation as a basic record type is a line segment .This storing pointers to the two end vertices and the two (anticlockwise) adjacent lin segments and triangulations are valid data structures whose relative advantages on the application |
| 6. Triangulation in this context expresses relationships between triples of objects, |
| 7. If triangulations express adjacency relationships between points (the duals triangulation is an appropriate expression of the adjacency relationships between is thus an expression of the adjacency relations between h original generating dat |
| 8. The objects associated with the triangle vertices need not be points : they ma points plus line segments |
| 9. The Voronoï criterion for any object is defined the same manner as for point calculated .Boundaries may be line segments or parabolas. |
| 10. The boundaries between Voronoï polygons are implicit in the relationship betw vertices (objects) in the triangulation ,and need not be preserved .The centre |

| | |
|---|---|
| junction between three Voronoï boundaries ) is more critical in determining which boundaries are to be preserved to form the triangulation . |
| 11. Basic operations for linked lists are: initialize; insert; delete; search; and switch. The equivalent for triangulations are : initialize (create a bounding triangle to enclose the data set) ; search (walk through the triangulation to find the bounding triangle for the pint or object ) ; insert ( split the bounding triangle into three to accommodate the new object ) ; switch ( interchange the diagonal on adjacent triangle pairs –performed if the Voronoï criterion is not met for the current triangle pair ) ; delete (remove an object from the triangulation by temporarily merging two adjacent objects and deleting the two redundant triangles : the reverse of insert ) |
| 12. The switch operation is performed whenever the common boundary between two adjacent triangles does not conform to the Voronoï criterion .For four points in isolation the Voronoï criterion guarantees that the one or other of the two ways of dividing the quadrilateral into triangles will be valid. |
| 13. The “insert” and “delete” operations are equivalent to “split” and “merge” operations on objects .This permits the hierarchical organizing of objects into a tree structure if required ,for efficient organization or searching .This is most easily understood if the dual of the objects ( a polygon set) is considered ; in this mode, two adjacent polygons A and B are merged into polygon AB ; |
| 14. Interpolation may be performed buy the judicious insertion and deletion of dummy sampling points in order to determine the relative areas of the adjacent Voronoï polygons stolen by the new dummy point. |
| 15. Line segments are constructed from their two end points and a connection link .If these end points and the line segments are inserted into the Voronoï network they will each generate their own Voronoï region . For line segments connected to form a polygon, the interior boundaries of these regions form the skeleton or media axis transform of the polygon in vector space |
| 16. Any triangulation may be processed as an oriented binary tree with respect to some viewpoint , permitting front-to-back or radially-outward ordering of objects on a business map .This is of use in contour construction , hidden line or surface removal , and the searching for all nearest neighbours within some tolerance . |

Table 1: The statements above form the conceptual stages in the design of an operational smart business network, in order to handle a variety of processes.